\documentstyle[12pt,epsf]{article}

\begin{document}
\begin{titlepage}
\begin{center}
\vspace{1cm}
\hfill
\vbox{
    \halign{#\hfil         \cr
         hep-th/9903108 \cr
           IPM/P-99/012\cr
           March 1999\cr} }
      
\vskip 1cm
{\Large \bf
Branes at Angles from DBI Action }
\vskip 0.5cm
{\bf R. Abbaspur\footnote{e-mail:abbaspur@netware2.ipm.ac.ir} 
}\\ 
\vskip .25in
{\em
Institute for Studies in Theoretical Physics and Mathematics,  \\
P.O. Box 19395-5531,  Tehran,  Iran.\\
Department of Physics,  Sharif University of Technology, \\
P. O. Box 19365-9161,  Tehran,  Iran.}
\end{center}
\vskip 0.5cm

\begin{abstract}

In this paper we investigate about several configurations of two intersecting 
branes at arbitrary angles. We choose the viewpoint of a brane source and a 
brane probe and use the low-energy dynamics of p-branes. For each p-brane this 
dynamics is governed by a generic DBI action including a WZ term, which 
couples to the SUGRA background of the other brane. The analysis naturally 
reveals two types of configurations: the ``marginal'' and the ``non-marginal'' 
ones. We specify possible configurations for a pair of similar or 
non-similar branes in either of these two categories. In particular, for two 
similar branes at angles, this analysis reveals that all the marginal 
configurations are specified by $SU(2)$ angles while the non-marginal 
configurations are specified by $Sp(2)$ angles. On the other hand, we find 
that no other configuration of two intersecting branes at non-trivial angles 
can be constructed out of flat p-branes. So in particular, two non-similar 
branes can only be found in an orthogonal configuration. In this case the 
intersection rules for either of the marginal or non-marginal configurations 
are derived, which thereby provide interpretations for the known results from 
supergravity.

\end{abstract}

\end{titlepage}\newpage

\def\ra{\rightarrow}
\def\={:=}
\def\p{\partial}

\def\a{\alpha}
\def\b{\beta}
\def\c{\gamma}
\def\d{\delta}
\def\e{\epsilon}
\def\ve{\varepsilon}
\def\g{\gamma}
\def\v{\upsilon}
\def\th{\theta}
\def\l{\lambda}
\def\m{\mu}
\def\n{\nu}
\def\o{\omega}
\def\x{\xi}
\def\r{\rho}
\def\s{\sigma}
\def\t{\tau}
\def\f{\phi}

\def\G{\Gamma}
\def\D{\Delta}
\def\L{\Lambda}
\def\O{\Omega}
\def\Th{\Theta}
\def\bO{\bf\Omega}

\def\cA{{\cal  A}}
\def\cB{{\cal  B}}
\def\cC{{\cal  C}}
\def\cF{{\cal  F}}
\def\cJ{{\cal  J}}
\def\cL{{\cal  L}}

\def\bi{{\bf 1}}
\def\bth{{\bf \Theta}}
\def\Db{{\bar D}}
\def\dt{{\tilde d}}

\def\st{sin\theta}
\def\sbt{sin^2\theta}
\def\ct{cos\theta}
\def\cbt{cos^2\theta}
\def\ea{\eta^\alpha}
\def\bX{{\bf X}}

\def\A{\parallel}
\def\B{\perp}
\def\ddd{\cdot\cdot\cdot}

\def\ft{\footnote}
\def\nn{\nonumber}
\def\be{\begin{equation}}
\def\ee{\end{equation}}
\def\bea{\begin{eqnarray}}
\def\eea{\end{eqnarray}}
\def\np{\newpage}

\section{Introduction}
Intersecting branes at arbitrary angles have been studied both in view
of their SUSY properties in supergravity
\cite{1,2,3,3*,4} and their short-range
interactions in string and M(atrix) theory \cite{3*,4,5}.
Despite the variety of the
information on SUSY properties of branes at angles, less
has been known about the structure of the
 supergravity solutions that describe 
 this type of configurations. The well known examples of these solutions
include the 2-angle (marginal) configuration of two
D2-branes at $SU(2)$ angles \cite{6}, 
 and the 4-angle (non-marginal) 
 configuration of two NS5-branes at $Sp(2)$ angles \cite{7}
and their generalizations to $n$ such branes \cite{6,7,8,9}
(For some other works in this respect see also \cite{7*,9a,9b}).
The solution for the (marginal) configuration  
at $SU(2)$ angles of two arbitrary similar branes
has been recently found in \cite{10}.
Nevertheless, there has been no
solutions that describe, for example, marginal intersections of
similar branes at three or more angles, or those for the non-similar branes
at other than the right angles. A systematic 
approach to the solution of this problem
can be found in the works of \cite{10,11}, where such solutions for a 
{\it distributed} system of branes at angles are considered. The analysis of
these solutions, in the light of a general formulation in \cite{11}, reveals
that, as far as we deal with flat p-branes, such solutions can not be realized
in terms of the harmonic functions. 
That is, in general, one can not  
find  configurations of branes at angles constructed from parallel
distributions of flat p-branes. 
The physics underlying this property
can be better understood, if one takes into account the role of the
brane(s) interactions and dynamics in forming a configuration at
several angles.
In fact, determining the stability conditions for 
any configuration of p-branes, based on the worldvolume  
dynamics of the branes, is logically
prior to any effort for finding a supergravity background that describes the
configuration.
The interactions of D-branes at angles have been studied 
by calculating the amplitudes in the scattering processes,
both in string and M(atrix) theory, 
 which determine the static  
potentials between pairs of these objects \cite{4,5}.
That such a potential {\it identically} vanishes determines the necessary 
condition for the stability of the configuration.
One of the basic ingredients
in such calculations, which makes them at all possible, is the assumption 
of {\it flatness} for the worldvolume  geometries of the individual branes 
as well as their spacetime background. However, there may
exist  configurations
for which the the two branes are not flat and so all features of their motion
can not be described by a single potential. For example, they may tend to
rotate relative to each other due to the `twisting forces' (or the
relative `torques') between themselves. So one has to add the conditions
that guaranty the balancing of such forces as well.
In fact, the full set of the stability conditions for a configuration of
two interacting branes are those that satisfy 
the  equations of motion for both of the branes when they are in {\it flat}
states.
Doing this, in general, involves extra complexities because of the
need for determining the supergravity background for an arbitrarily
curved p-brane.
For flat BPS branes, however, this problem is considerably simplified by the
fact that a BPS brane in this state
does not suffer from `self-interaction' forces.
Indeed, in such a case the forces on each individual brane, due to coupling
to its own supergravity background, are balanced against each other and
hence, it can be studied as a {\it probe} scattered from another brane,
which is the {\it source} of a well known supergravity background \cite{14}.
The aim of this paper is to study and classify possible 
configurations of branes at angles,
assuming the DBI+WZ worldvolume action as the branes dynamics. 
To this end, we first determine the general necessary
and sufficient conditions that specify a stable configuration of two arbitrary
flat branes in section 2. 
Then, in sections 3,4,5, we use these general conditions for
categorizing several configurations of two branes at angles for the cases of
similar, non-similar and electromagnetic dual branes. We will consider both
the marginal and non-marginal bound states in these sections.
Multi-angle marginal intersections are discussed in section 5, 
and at the end, we derive
physical interpretations for our stability conditions in section 7.
We end the paper by a summary and some remarks.

\section{General set up for the stability conditions}
As the discussion in the introduction implies, 
in all cases with flat worldvolume
geometries, one can describe a two brane system, equivalently, by the
worldvolume action for each of its constituent branes.
For simplicity, we assume that $d_1\leq d_2$ and we take the $(d_1-1)$-brane
as a probe moving in the background fields produced by a $(d_2-1)$-brane
source.
To describe this configuration, we shall use an orthogonal coordinate
system whose axes are defined using the tangent and normal directions
of the two branes as indicared in table (1).
 In this table
$d \= d_1+d_2-\d$ stands for the  dimension of the hyperplane spanned
by the world directions of the two branes, and $D$ denotes the spacetime 
dimension. The situation is  schematically displayed  in figure (1).
$$
\begin{array}{|c|c|l|}\hline
coordinates&dimensions&\ \ \ \ definitions\\ \hline
x^{\r}&\d&x^{\r}\A d_1\ ,\ x^{\r}\A d_2 \\ \hline
x^{r}&(d_1-\d )&x^{r}\B \d\ ,\ x^r\B y^i\ ,\ x^{r}\A d_2 \\ \hline
y^i&(d_2-d_1)&y^i\B d_1\ ,\ y^i \A d_2\\ \hline
y^r&(d_1-\d )&y^r\A d \ , \ y^r \B d_2 \\ \hline
z^a&(D-d)&z^a\B d_1\ ,\ z^a\B d_2 \\ \hline

\end{array}
$$                               
\begin{center} Table (1):
decomposition of the spacetime coordinates  \end{center}

\begin{figure}[t]
\begin{center}
%\leavevmode
\vspace{-30mm}
\hspace{5mm}
\epsfxsize=150mm
\epsfysize=100mm
\epsfbox{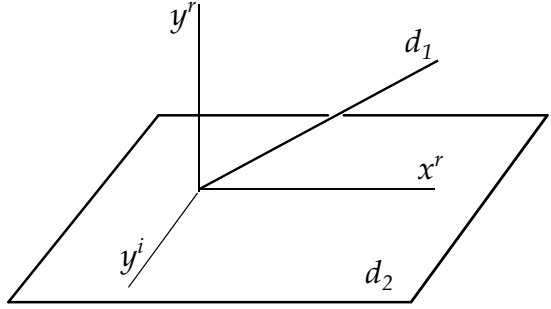}
\caption{{\it representation of the branes worldvolumes and the 
 spacetime coordinates
(note that coordinates $x^\r$ and $z^a$ can not be displayed in this figure) }}
%\label{fig:?}
\end{center}
\end{figure}

We choose $x^{\a}\= (x^{\r},x^r)$ as the parameterizing coordinates of the
probe and $y^A\= (y^i,y^r,z^a)$ as its embedding coordinates.
According to the count of coordinates in table (1), there are cases  
for which some of the coordinates in our decompoition are not present. 
For example, when $d_1=d_2$, we have no $y^i$
coordinates. Similarly, when $d_1=\d$ (i.e., the two branes are parallel),
there are no $x^r$ and $y^r$ coordinates. However, this has no effect on
our general equations as they are formulated in terms of the coordinates
 $x^{\a}$ and $y^A$ regardless their numbers. 

Thus, the embedding coordinates for a probe 
rotated/boosted relative to a fixed
source are represented as
\be
Y^A(x)={\o}^A_{\a}x^{\a}+y^A_0
\label{1}\ee
with $\o$'s and $y_0$'s representing constant slopes/velocities and shifts.
It is clear from the definitions in table (1) that
\be
{\o}^A_{\r}={\o}^i_{\a}={\o}^a_{\a}=0
\label{2}\ee
That is, $(Y^i,Z^a)$ are constants and none of $Y^A$'s depend on $(x^{\r})$.
Evidently, the remaining ${\o}$'s can not have arbitrary values. In fact,
they have to be chosen in a way that eq.(\ref{1}) solves the $(d_1-1)$-brane
equations of motion:
\be
{\p}_{\a}\left ({{\p\cL}\over{\p Y^A_{,\a}}}\right )-{{\p \cL}\over{\p Y^A}}=0
\label{3}\ee
where $\cL (Y^A,{\p}_{\a}Y^A)$ denotes the relevant worldvolume Lagrangian.
That is, $\o$'s must be such the identities
\be
{{\p}\over{\p Y^B}}\left ( {\o}^B_{\a}{{\p\cL}\over{\p {\o}^A_{\a}}}
-{\d}^B_A\cL \right )\equiv 0
\label{4}\ee
hold for all values of $(x^r)$.
Here, $\cL$ have to be considered as the function $\cL (Y^A;{\o}^A_{\a})$,
whose $Y^A$-dependences turn to appear in the form 
\be
\cL (Y^A;{\o}^A_{\a})=\cL\left (H(\sqrt{(Y^r)^2+(Z^a)^2})
\ ;\ {\o}^A_{\a}\right )
\label{5}\ee
where $H$ represents a harmonic function in the
$(y^r,z^a)$-directions \cite{12}. Taking this into account,
 and that $Y^r$'s are
independent coordinates as far as $x^r$'s are so,
one can see that eq.(\ref{4}) leads to
\be
{{\p\cL}\over{\p H}}Z^a=0
\label{6}\ee
\be
{{\p}\over{\p H}}\left ({\o}^r_{\a}{{\p\cL}\over{\p{\o}^s_{\a}}}
-{\d}^r_s\cL\right )=0
\label{7}\ee
We will refer to eqs.(\ref{6}),(\ref{7}) as the `no-force' and `no-torque' 
conditions, respectively, since they are indeed the conditions for the
vanishing of the oscillations along the transverse
directions of the two branes, as well as
 their (relative) rotations along the relative transverse directions
(see section 7).

\subsection{The marginal and non-marginal configurations}
By definition, a `marginal' configuration of two branes is the one
which is stable at any arbitrary separations of the 
two branes, i.e., for any {\it constant}
values of $Z^a$'s. This, according to eq.(\ref{6}), implies
\be
\cL (H;\o )\equiv {\cL}_0(\o )
\label{8}\ee
showing that the total potential energy between the two branes
has a constant value. It is obvious that eq.(\ref{7}) in this case is
automatic and hence the above equation is the unique description of the
`marginal' configurations.\\

There exist other configurations, however, that can be formed only at zero
separations of the two participating branes, i.e., when the centers
of the two branes in their transverse space are coincident and they actually
intersect each other. We refer to these as the non-marginal configurations
\cite{13}.
For such solutions, with $Z^a=0$, eq.(\ref{6}) is automatic and hence 
eq.(\ref{7}) or its equivalent as
\be
{\o}^r_{\a}{{\p\cL}\over{\p {\o}^s_{\a}}}-{\d}^r_s\cL\equiv {\cC}^r_s (\o )
\label{9}\ee
provides a unique description for the non-marginal configurations.
In the following three sections, we analyze the eqs.(\ref{8}),(\ref{9})
for classifying several marginal and non-marginal configurations that may
occur in cases with a pair of similar $(d_1=d_2)$, non-similar $(d_1\neq d_2)$
and electromagnetic dual $(d_1={\dt}_2)$ branes. 
The notations will be mainly
identical to those introduced in \cite{10,12}.

\section{Configurations with similar branes}
Using the general expression of the Lagrangian for a $(d-1)$-probe in a 
$(d-1)$-source background \cite{12}, the function $\cL (H;\o)$ is found to be
\be
\cL =H^{-1}[det^{1/2}(\bi +H\bth )-1]
\label{10}\ee
where the angular information is encoded in the matrix $\bth$ having the 
components
\be
{\Th}^{\a}_{\b}\= {\o}^{r\a}{\o}^r_{\b}
\label{11}\ee
and the indices of the type of $\a$ are raised and lowered by 
${\eta}_{\a\b}\= diag(-1,+1,\ddd ,+1)$.
The matrix $\bth$ can be diagonalized by choosing the coordinates 
$(x^r,y^r)$ so that the two branes to be related by a set of `commuting'
rotations and a boost. In such coordinates, $\bth$ has the components
\be
{\Th}^{\a}_{\b}=diag(-v^2,tan^2{\th}_1,\ddd ,tan^2{\th}_{d-1})=:
{\d}^{\a}_{\b}{\Th}_{\a}
\label{12}\ee
Here, $v$ represents the magnitude of the velocity vector in the directions 
transverse to the $(d-1)$-source and $\th$'s are the angles of the $(d-1)$
commuting rotations (obviously, a number $\d$ of these angles are vanishing
by assumption). 

\subsection{Marginal intersections}
Inserting the expression (\ref{10})
for $\cL$ into the `marginality' condition, eq.(\ref{8}), one obtains
\be
det(\bi +H\bth )\equiv (1+{\cL}_0H)^2\ \ \ \ \ (\forall H>0)
\label{13}\ee
In the basis that $\bth$ has the form of eq.(\ref{12}),
this identity reads
\be
(1-v^2H)(1+tan^2{\th}_1H)\ddd (1+tan^2{\th}_{d-1}H)\equiv (1+{\cL}_0H)^2
\label{14}\ee
Thus the only possibility for the rotation and boost parameters is 
\bea
&&{\th}_1=\pm{\th}_2=:\th\nn\\   
&&v={\th}_3=\ddd ={\th}_{d-1}=0
\label{15}\eea
That is: \\
the only marginal configuration of two similar 
branes are the static configurations with two non-vanishing angles obtained 
by an abelian subgroup of the $SU(2)$ rotations.\\
This is the same as the 1/4 SUSY configuration whose supergravity solution
had been found earlier in \cite{10} as a generalization
to the solution of 2-branes at angles \cite{6}.
The constant value of
$\cL$ in this case is ${\cL}_0=tan^2\th$. \\
Another possibility, which one might consider for two similar branes, was the 
brane-anti-brane system. In this case the constant term $(-1)$ in brackets 
in eq.(\ref{10}) would be flipped to (+1) which is equivalent to flipping
${\cL}_0$ to $-{\cL}_0$ in eq.(\ref{13}) or (\ref{14}). 
However, because ${\cL}_0\geq 1$ in this case, eq.(\ref{14}) could not
be satisfied for any real values of $v$ and $\th$'s. In other words, a 
brane-anti-brane pair with any combination of a boost and several rotations 
is unstable and can not form a marginal configuration.

\subsection{Non-marginal configurations}
It is easy to see that the `non-marginality' condition, eq.(\ref{9}),
after some algebra reduces to
\be
\sqrt{\O}({\bO}^{-1})^{\a\b}{\o}^r_{\a}{\o}^s_{\b}-
{\d}^{rs}H^{-1}(\sqrt{\O}-1)={\cC}^{rs}
\label{16}\ee
where ${\bO}\={\bi}+H\bth$ , $\O\=det\bO$ and $({\cC}^{rs})$ are a set of
angle-dependent constants. In the basis that makes $\bth$ diagonal
(eq.(\ref{12})), the above equation is also diagonal and after some
manipulations yields 
\be
\prod_{\a}(1+H{\Th}_{\a})\equiv (1+H{\Th}_r)^2(1+H{\cC}_r)^2
\ \ \ \ \ \ (\forall H>0)
\label{17}\ee
where ${\cC}^{rs}=:{\d}^{rs}{\cC}_r$. For this equation to be satisfied
identically, one needs
\bea
&&{\th}_1=\pm{\th}_3\ ,\ {\th}_2=\pm{\th}_4\nn\\
&&v={\th}_5=\ddd ={\th}_{d-1}=0
\label{18}\eea
That is:\\
the only `non-marginal' configurations of two `similar' branes are the static
configurations with four angles obtained by two independent abelian subgroup
of the $SU(2)$ rotations.\\

It is easy to  check that, unless one of the $({\th}_1,{\th}_2)$ vanishes,
in this case $\cL$ is not a constant, but it is a linear function of $H$.
Orthogonal configurations of the
4-angle intersections have been found previously
in the literature of the supergravity composite brane solutions \cite{16}. The
most famous examples in this category consist of the
$NS5\cap NS5=1$ in $D=10$ II A,B ,  $D5\cap D5 =1$ in $D=10$ II B
and $M5\cap M5=1$ in D=11 \cite{16}. In a recent paper \cite{7},
a 4-angle configuration of this type for
the II A NS5-branes  at Sp(2) angles
has been found, by directly solving the supergravity and Killing spinor
equations, showing that it preserves at least 3/32 of the SUSY
(see however \cite{7*,2,3,4}).

\section{Configurations of two non-similar branes}
The general worldvolume Lagrangian for a $(d_1-1)$-probe in a
$(d_2-1)$-source background, with $d_1<d_2$ \cite{12}, gives rise to
an expression for $\cL (H;\o)$ as
\be
\cL =H^{-m/2}det^{1/2}(\bi +H\bth)
\label{19}\ee
where $\bth$ is given by eq.(\ref{11}) and $m$ is a function of
dimensions defined as
\be
m(d_1,d_2)\= 2\a (d_1)\a (d_2)+{{d_1{\dt}_2}\over{D-2}}
\label{20}\ee
For marginal configurations, $m$ specifies the number of (non-vanishing)
angles (see \cite{10} and below). Here, $\a (d)$ represents the dilaton-d-form
coupling constant satisfying \cite{14}
\be
{\a}^2(d)=1-{{d\dt}\over{2(D-2)}}
\label{21}\ee

\subsection{Marginal intersections}
The marginality condition, eq.(\ref{8}), in this case gives
\be
\prod_\a (1+H{\Th}_{\a})\equiv {\cL}_0^2H^m\ \ \ \ \ \ \ (\forall H>0)
\label{22}\ee
Obviously, such an identity can be true whenever $m\in  {\bf Z}^+$, and
a number $m$ of ${\Th}_{\a}$'s to be $+\infty$, while the remaining 
are vanishing. This implies that
\bea
&&{\th}_1=\ddd ={\th}_m=\pi /2\nn\\
&&{\th}_{m+1}=\ddd ={\th}_{d_1-1}=v=0
\label{23}\eea
The number of common directions then is $\d =d_1-m$. Therefore:\\
the only `marginal' bound state of a pair of `non-similar'
$(d_1-1,d_2-1)$-branes are the `static orthogonal' configurations in which
the two branes share $(\d -1)$ of their directions, where $\d$ must satisfy
\be
\d =-2\a (d_1)\a (d_2)+{{d_1d_2}\over{D-2}}
\label{24}\ee
This is, in fact, the so called `intersection rule' of the intersecting
brane systems, which was originally found in the study of their supergravity
solutions \cite{15,15*},
 and then re-appeared in the reduced Lagrangian approach
to the distributed brane systems in \cite{10} and was interpreted there
as an algebraic constraint 
\footnote{In fact, eq.(\ref{24}) can be viewed as  a
relation between the gravitational and dilatonic
`charges' $(M_{\l},{\a}_{\l} M_{\l})$, $\l =1,2$ of the two branes, with
$M_{\l}$ being their masses,
for the corresponding forces to cancel each other \cite{10}.}
required by the no-force conditions \cite{18}.

\subsection{Non-marginal Intersections}
The `non-marginality' condition, eq.(\ref{9}), in this case gives the
counterpart of eq.(\ref{16}) as
\be
H^{-m/2}\sqrt{\O}\{ H({\bO}^{-1})^{\a\b}{\o}^r_{\a}{\o}^s_{\b}-
{\d}^{rs}\} ={\cC}^{rs}
\label{25}\ee
which in the $\bth$-diagonalizing basis (eq.(\ref{12}))
takes the form of the identity
\be
\prod_{\a}(1+H{\Th}_a)\equiv {\cC}_r^2H^m(1+H{\Th}_r)^2\ \ \ \ \ \ \ \ \
(\forall H>0)
\label{26}\ee
which is the analogue of eq.(\ref{17}). Obviously, such an identity holds
(for each $r$), if and only if $m\in {\bf Z}^+$, and further $(m+2)$ of
${\Th}_{\a}$'s are infinity while the others are vanishing. Therefore,
\bea
&&{\th}_1=\ddd ={\th}_{m+2}=\pi /2 \nn\\
&&{\th}_{m+3}=\ddd ={\th}_{d_1-1}=v=0
\label{27}\eea
These are just the conditions (\ref{23}) in which $m$ 
has been replaced by $(m+2)$.
So we have the result:\\
the only `non-marginal' bound state of a pair of `non-similar'
$(d_1-1,d_2-1)$-branes are the `static orthogonal' configurations in which
the two branes share $(\d -1)$ of their directions, with $\d$ satisfying
\be
\d +2=-2\a (d_1)\a (d_2)+{{d_1d_2}\over{D-2}}
\label{28}\ee
which shifts $\d$ by a $-2$ relative to the one given by the rule (\ref{24}).
This is just the same equation identified as the intersection rule
of the `localized intersections' in \cite{16}. By the above derivation, however,
it has to be identified as a rule for the `non-marginal' intersections.

\section{Configurations with an electromagnetic dual pair of branes}
An `electromagnetic' dual pair of branes, a priori, can not be placed in
either of the two categories studied in sections 3,4. This is due
to the fact that an electromagnetic dual pair with $(d-1,\dt -1)$-branes
is the source of a single $d$-form potential ${\cA}_{(d)}$, which couples to
the two branes through ${\cF}_{(d+1)}=d{\cA}_{(d)}$ and its dual
$^*{\cF}_{(d+1)}$ respectively \cite{14}. Taking this point into account,
and assuming $d\leq \dt$ to adapt with the conventions of section 2, one
obtains the modified version of eq.(\ref{4}) as
\be
{{\p}\over{\p Y^n}}\left ( {\o}^n_{\a}{{\p\cL}\over{\p {\o}^m_{\a}}}
-{\d}_{mn}\cL -{\c}_{mn}H \right )\equiv 0
\label{29}\ee
where $\cL$ is defined in eq.(\ref{19}) in which $m=d-2$, and the
constants ${\c}_{mn}$ are defined as
\be
{\c}_{mn}\= {\e}_{mnl_1\ddd l_d}{\e}^{{\a}_1\ddd {\a}_d}
{\o}^{l_1}_{{\a}_1}\ddd {\o}^{l_d}_{{\a}_d}/(d!)^2
\label{30}\ee
Here, the two epsilons stand for the Levi-Civita symbols in the subspaces
of $(y^m)\= (y^r,z^a)$ and $(x^{\a})$, respectively. The modifying term
${\c}_{mn}H$ in eq.(\ref{29}), however, does {\it not} modify the results
of section 4 for the branes at angles with $(d_1,d_2)=(d,\dt )$.
The reason, as can be checked using eq.(\ref{12}), is that in all $\d \geq 1$
cases one obtains ${\c}_{mn}=0$. Thus the rule (\ref{24})
(with $\a (d)=-\a (\dt )$) and eq.(\ref{21}) imply $\d =2$. Hence:\\
the only marginal configuration of an electromagnetic dual pair of
$(d-1,\dt -1)$-branes is a `static' configuration in which the
two branes `orthogonally' intersect (overlap) on a string.

On the other hand, a non-marginal configuration should obey the
rule (\ref{28}), which in this case implies that $\d =0$. This is in
obvious contradiction with the static-ness property asserted above the rule
(\ref{28}) which means that such configurations of $(d,\dt )$ can not be
realized.

\subsection{Non-marginal configuration of parallel branes}
 
The results of the previous sections regarding the non-marginal configurations
, in particular their intersection rule (eq.(\ref{28})), are based on the 
no-torque condition (eq.(\ref{7})) which  severly relies on the existence of
at least one pair of coordinates $(x^r,y^r)$ in table (1). 
Since for a pair of parallel 
branes (i.e. a configuration of the form $d_1\subseteq d_2$) there are no
such coordinates, hence the stability condition for such configurations should 
have
a different form. Whatever the form of these conditons, they should 
guarantee that the equations of motion (eq.(\ref{4})) are satisfied. In the 
case at hand all ${\o}^A_\a$'s are zero and this equation reduces to:  
$\p\cL /\p Y^A=0$, where $\cL (Y^A)$ is the same as $\cL \left (H(Y^A);
{\o}^A_{\a}=0\right )$. Noting that $H$ in this case depends only on $Z^a$
through $r:=\sqrt{Z^a Z^a}$, and further that $Z^a=0$ in the non-marginal
case, the above equation yields:
\be
{{\p V(r)}\over {\p r}}\large |_{r=0}=0
\label{*}
\ee
where $V(r):=\cL \left (H(r);{\o}^A_{\a}=0\right )$. For a pair of 
non-similar  branes we find from eq.(\ref{19}) that
\be
V(r)=(H(r))^{-m/2}=\left (1+{{Q_2}\over {r^{M-2}}}\right )^{-m/2}
\ee
Here $M:=\tilde d_2+2$ is the transverse dimension and $Q_2$ is (proportional
to) the charge of $(d_2-1)$-brane and $m$ is defined as in eq.(\ref{20}). 
(We have assumed that $d_1\leq d_2$ and $\tilde d_2>0$.)
It is 
obvious that $V(r)$ is proportional to the potential energy between the two 
branes at a separation $r$ and eq.(\ref{*})  indicates 
that their mutual forces tend to balance each other as $r\ra 0$. 
Now, since in this 
limit $V(r)\sim r^{m(M-2)/2}$, the above condition means that the function
$V(r)$ should be regular near $r=0$ at least to the first order in its 
derivatives, requiring that
\be
m(M-2)>2 
\ee
or equivalently
\be
{\a}_1{\a}_2>{{2(D-2) -d_1\tilde d_2^2}\over{2\tilde d_2(D-2)}}
\ee
This inequality together with $d_1\leq d_2$ and $\tilde d_2>0$ in turn
specifies all the non-marginal bound states of the form of branes within
branes. Special cases of such configurations are those with a self dual 
pair, i.e. with  $d_1=d \ ,\ d_2=\tilde d$. Since in this case $M=d+2 $ , 
$m=d-2$, the above condition (assuming $d\leq\tilde d$) yields: 
$3\leq d\leq (D-2)/2$ showing that such configurations are possible only
in $D\geq 8$ dimensions. Famous examples of such configurations are 
$M2\subset M5$ in $D=11$ and the dyonic membrane 
(bound state of an electric
and a magnetic 2-brane) in $D=8$ dimensions, which were known through their 
 supergravity solutions \cite{17}.

\section{Multi-angle marginal intersections}
The results of the previous sections indicate that, except for the
2-angle static configuration of similar branes at $SU(2)$ angles, no other
configuration of two flat branes with several angles and boost can be
marginally stable. While, both the scattering amplitude and superalgebra
computations \cite{4,2} indicate that multi-angle intersections, under certain
conditions among their angles, can form a marginal configuration, it is
amazing that the worldvolume solutions appropriate to
a pair of flat branes are not realizable. 
 This apparent contradiction is resolved by recalling that
all the scattering amplitude and superalgebra computations rely
heavily on the basic assumption for the existence of 
an `asymptotic state' in which the branes 
behave like flat hypersurfaces in a Minkowski space. 
In the string theory language, the perturbative calculation of the 
scattering amplitudes are reliable only in the weak coupling region of the 
theory where one deals with weak (linearized) gravitational interactions. 
This restricts such calculations to the `far' or `asymptotic' region of the
two branes where they look like flat hypersurfaces in a flat spacetime.
On the SUSY side, also, one does not need to solve a Killing spinor equation 
in all of the space to determine SUSY fraction preserved by a brane 
configuration. To do this, it suffices to solve only the asymptotic 
(algebraic) Killing spinor equation \cite{1,2,3}, which encodes {\it only} the
asymptotic form of the spacetime metric and of the geometry of branes.
As a result, there may be BPS states of {\it curved} p-branes
which look like asymptotically as multi-angle configurations of flat p-branes.
In the asymptotic region, we will see that the no-force condition to 
first order will require a certain angular constraint which is just the same
that characterizes these BPS states \cite{4,5}. However, the same no-force  
condition to higher orders, as well as the no-torque condition, 
are not satisfied except for the marginal configurations which were 
categorized in sections 3,4. 
This means, firstly, that the marginal multi-angle configurations, 
generally, can not consist of flat p-branes. Secondly, even in the case of an
asymptotically force-free configuration, the relative angular position of the 
two branes is influenced by a non-vanishing torque which eventually
brings them together by counterbalancing the forces that act 
between themselves.
In general, the force and torque conditions, eq.(\ref{8}),(\ref{9}),  
are equivalent to a set of algebraic constraints relating ${\Th}^{\a}_{\b}$'s
together. This can be seen easily by putting $H\= 1+h$ and expanding 
$\cL (H;\o )$ as
\be
\cL (1+h;\o )=\sum_{n=0}^{\infty}{{1}\over{n!}}{\cL}_n(\o )h^n
\label{32}\ee
Upon this expansion, eqs.(\ref{8}),(\ref{9}) give respectively
\be
{\cL}_n(\o )=0
\label{33}\ee
\be
{\o}^r_{\a}{{\p{\cL}_n}\over{\p{\o}^s_{\a}}}-{\d}^r_s{\cL}_n=0
\label{34}\ee
where $n=1,2,\ddd $ . For a `real' marginal bound state of flat p-branes,
these two sets of equations for all $n$ restrict possible configurations  
to those of sections 3,4. However, for a configuration of curved p-branes
with asymptotic flat geometries, we may continue to {\it define}
the marginality property by demanding that $\cL (H;\o )$ to be constant 
{\it only} to first order in $(H-1)$ in the asymptotic region 
$H(y)\ra 1$. Such `asymptotic marginal' configurations, thus, are
distinguished by a condition on the angles as ${\cL}_1(\o )=0$. 
This condition, though provides a mean of translational stability, does
{\it not} insure rotational stability of the two brane system, which to 
be guaranteed by the eq.(\ref{34}) for $n=1$. These two conditions together,
will be seen that, restrict the possible marginal configurations to those 
obtained in sections 3,4. We now examine the explicit expressions 
of these conditions in the previous cases.

\subsection{Similar branes}
It is easy to see, by expanding eq.(\ref{10}) to ${\cal O}(h)$, that in 
this case 
\bea
&&{\cL}_0(\Th )=det^{1/2}(\bi +\bth )-1\nn\\
&&{\cL}_1(\Th )=det^{1/2}(\bi +\bth )\left \{ {1\over 2} Tr\left (
{{\bth}\over{\bi +\bth}}\right )-1\right \}+1
\label{35}\eea
where $\o$ dependences are encoded in $\bth$ as is defined in eq.(\ref{11}).
(The expression for ${\cL}_0$ is given only for later reference.)
Diagonalizing $\bth$ as in eq.(\ref{12}), the ${\cL}_1=0$ equation becomes
\be
F(\th )\= \prod_{\a =0}^{d-1}cos{\th}_{\a}+
{1\over 2}\sum_{\a =0}^{d-1}sin^2{\th}_{\a}-1=0
\label{36}\ee
where, for convenience, we have included the velocity $v$ in $\th$'s by
defining ${\th}_0\= tan^{-1}(iv)$.
Solving eq.(\ref{36}) for $v$, however, shows that combinations of boost  
and rotations for less than three non-vanishing angles do {\it not}
define allowable configurations. Indeed, one can see using eq.(\ref{36}),
that the one- and two-angle configurations are limited only to the parallel 
and $SU(2)$-rotated static configurations respectively.
If, in addition to the above condition, one requires rotational
stability of the configuration, eq.(\ref{34}) for $n=1$ gives
\be
{{\p F}\over{\p {\th}_{\a}}}=sin{\th}_{\a}\left (\prod_{\b\neq\a}cos{\th}_{\b}
-cos{\th}_{\a}\right )=0
\label{37}\ee
The only simultaneous solutions of eqs.(\ref{36}),(\ref{37}), when more than
two ${\th}_{\a}$'s exist, are the one with
\be
{\th}_1={\th}_2 \ \ ,\ \ {\th}_0={\th}_3=\ddd ={\th}_{d-1}=0
\label{38}\ee
and its permutations for ${\th}_{\a}$'s ($\a\neq 0$). That is, the only 
rotationally stable marginal configurations are those with $SU(2)$ angles.

\subsection{Non-similar branes}
In this case eq.(\ref{19}) gives 
\bea
&&{\cL}_0(\Th )=det^{1/2}(\bi +\bth )\nn\\
&&{\cL}_1(\Th )={1\over 2}det^{1/2}(\bi +\bth )\left \{  Tr\left (
{{\bth}\over{\bi +\bth}}\right )-m\right \}
\label{39}\eea
So, in the basis of eq.(\ref{12}), the ${\cL}_1=0$ condition becomes
\bea
&&F(\th )\= \sum_{\a =0}^{d_1-1}sin^2{\th}_{\a}-m=0\ \ \ \Rightarrow \nn\\
&&-2\a (d_1)\a (d_2)+{{d_1d_2}\over{D-2}}=\sum_{\a =0}^{d_1-1}cos^2{\th}_{\a}
\label{40}\eea
This gives, in fact, a modification of the usual intersection rule, 
eq.(\ref{24}), to the general case involving arbitrary boost and angles
between the two branes. Obviously, an `orthogonal  static' limit, with 
${\th}_{\a}$'s equal to $0,\pi /2$, 
exists only in cases with $m\in {\bf Z}^+$.
That is, for an orthogonal intersection, the rule (\ref{24}) must hold and in
such a case $m$ counts the number of ${\th}_{\a}=\pi /2$ angles.
Despite eq.(\ref{36}), the eq.(\ref{40}) allows for the possibility of the
combinations of a boost with any number of angles. 
Specially, when $m\leq 0$, one can find (asymptotically) boosted 
configurations of two parallel branes having a relative velocity
$v=\sqrt {{m\over{m-1}}}$.
However, configurations defined by eq.(\ref{40}) are not rotationally stable,
unless we have
\be
{{\p F}\over{\p {\th}_{\a}}}=2sin{\th}_{\a}cos{\th}_{\a}=0
\label{41}\ee
which means that ${\th}_{\a}$'s must be $0,\pi /2$. Therefore, the
rotationally stable marginal configurations are the static ones with
orthogonal branes obeying the rule (\ref{24}).\np

\section{Small perturbations on the worldvolume}
So far, we have stressed on the fact that every `equilibrium'
configuration of two flat p-branes is specified by the two requirements,
eqs.(\ref{6}),(\ref{7}), which we have interpreted as the no-force and
no-torque conditions respectively. In this section, we try to make this
correspondences explicit by perturbing the worldvolume of one of the branes
(the probe) with respect to its flat state, while the other brane (the source)
is kept flat and fixed. This type of description is similar to the one which
is used in \cite{12}. To be concrete,  we first present a perturbative
formulation of the classical solutions for a general field theory
with one expansion parameter,  and then
consider its application to the worldvolume field theory of the probe.

\subsection{General field theory formulation}
Assume that a field theory, for the variable(s) $\f (x)$,
is defined by means of the perturbative Lagrangian
\be
\cL [\f ;\ve ]={\cL}_0[\f ]+\ve{\cL}_1[\f ]+
{{{\ve}^2}\over 2}{\cL}_2[\f ]+\ddd 
\label{42}\ee
with $\ve$ being the perturbation parameter.
Now, take ${\f}_0(x)$ to represent
a classical solution of the unperturbed Lagrangian ${\cL}_0[\f ]$, i.e.,
it satisfies the equation of motion: ${{\d {\cL}_0}\over{\d\f}}=0$. The
question, which we like to answer, is that how can we specify a solution of
the perturbed Lagrangian which tends to ${\f}_0(x)$ when $\ve\ra 0$.
Obviously, such a solution must obey the expansion
\be
\f (x;\ve )={\f}_0(x)+\ve {\f}_1(x)+{{{\ve}^2}\over 2}{\f}_2(x)+\ddd 
\label{43}\ee
Putting this expansion in eq.(\ref{42}), and using the functional Taylor
series expansion of ${\cL}_n$'s, the overall expansion of
$\cL [\f (x;\ve );\ve ]$ in powers of $\ve$ takes the form
\footnote{In fact, such an equation is only a symbolic expression in which the
functional derivatives of ${\cL}_n$'s are in the form of local operators,
acting on the functions which are multiplied by themselves,
summed over the field indices and are finally
integrated over the spacetime coordinates.}
\be
\cL [{\f}_0,{\f}_1,\ddd ;\ve ]={\cL}_0[{\f}_0 ]+
\ve \left ({{\d{\cL}_0}\over{\d\f}}{\f}_1+{\cL}_1\right )+
{{{\ve}^2}\over 2}\left ({{\d{\cL}_0}\over{\d\f}}{\f}_2+
{{{\d}^2{\cL}_0}\over{\d{\f}^2}}{\f}_1^2+2{{\d{\cL}_1}\over{\d\f}}{\f}_1+
{\cL}_2\right )+\ddd 
\label{44}\ee
where all ${\cL}_n$'s and their functional derivatives are evaluated
at $\f ={\f}_0$. Treating $\{{\f}_n\}$ as a set of independent variables,
and varying $\cL$ with respect to these variables, one picks the set of
equations
\bea
&&{{\d{\cL}_0}\over{\d\f}}=0\nn\\
&&{{{\d}^2{\cL}_0}\over{\d{\f}^2}}{\f}_1+{{\d{\cL}_1}\over{\d\f}}=0\nn\\
&&{{{\d}^2{\cL}_0}\over{\d{\f}^2}}{\f}_2+
{{{\d}^3{\cL}_0}\over{\d{\f}^3}}{\f}_1^2+
2{{{\d}^2{\cL}_1}\over{\d{\f}^2}}{\f}_1+
{{\d{\cL}_2}\over{\d\f}}=0 \ ,\ etc.
\label{45}\eea
These equations, in principle , determine the solutions for ${\f}_0(x),
\ {\f}_1(x),\ etc.$ in the successive order. It is worth-pointing that,
except for ${\f}_0(x)$ which is assumed to be a given solution of ${\cL}_0$,
all other ${\f}_n$'s are determined by solving
a linear inhomogeneous PDE of the form
\be
{{{\d}^2{\cL}_0}\over{\d{\f}^2}}{\f}_n={\cJ}_n
\label{46}\ee
where the source term ${\cJ}_n(x)$ in 
the $n$-th stage is a known  function of $x$,
constructed from ${\f}_0(x),\ddd ,{\f}_{n-1}(x)$. For vanishing boundary
conditions on ${\f}_n$'s, one solves eq.(\ref{46}) symbolically as
${\f}_n=G{\cJ}_n$, with $G$ representing the Green's function of the linear
operator ${{{\d}^2{\cL}_0}\over{\d{\f}^2}}$ at $\f ={\f}_0$.
If ${\cL}_0$, as in usual, is a first order Lagrangian in terms of the
$\f$-derivatives, then ${{{\d}^2{\cL}_0}\over{\d{\f}^2}}$ will be a second
order linear differential operator with (in general) spacetime dependent
coefficients, whose Green's function is constructed in the usual manner.
The above procedure, thus eventually, determines $\f (x;\ve )$ to any
arbitrary order in the expansion parameter $\ve$.

\subsection{Application to the worldvolume field theory}
We consider a source-probe configuration of arbitrary branes, as in
\cite{12}, and choose the embedding of the probe, as in section 2,
to be represented by $Y^A=Y^A(x^{\a})$. It is clear that the
worldvolume  Lagrangian of the probe has a generic form as
\be
\cL [Y]=\cL \left (h(Y^A),{\p}_{\a}Y^A\right )
\label{47}\ee
where the explicit dependences on $Y^A$ are encoded in the harmonic
function $h$ of the transverse distance from the source, which is proportional
to its charge or tension $T_s$, and vanishes asymptotically.  Obviously,
taking the limit $T_s\ra 0$ is equivalent to going to the asymptotic region
of $Y^A$, where the source and probe have a large separation, and in this
limit eq.(\ref{47}) takes the form of an ordinary Lagrangian of a minimal
surface in the Minkowski space. Taking this as the unperturbed Lagrangian
${\cL}_0$, and treating $T_s$ as a perturbation parameter, the perturbed
Lagrangian $\cL$ will be expanded as
\be
\cL [Y]=\sum_{n=0}^{\infty}{1\over{n!}}h^n(Y){\cL}_n(\p Y)
\label{48}\ee
which is nothing but the eq.(\ref{32}) with ${\o}^A_{\a}$'s replaced by
${\p}_{\a}Y^A$'s. Also, it is clear that ${\cL}_0$ has classical solutions
which are in the form of flat hypersurfaces, similar to the one
in eq.(\ref{1}), which plays here the role of the `unperturbed solution'
$Y^A_0(x)$. The main advantage of this choice for $Y^A_0(x)$ is that it
renders ${\cL}_n[Y_0]$'s as constant parameters
${\cL}_n(\o )$. This means that we are
interested in a probe whose worldvolume geometry, in the region far from
the source, is that of a flat hypersurfaces, though it may be curved in the
near region. Thus, perturbing the flat solutions as $Y^A(x)=Y^A_0(x)+
Y^A_1(x)+\ddd $, and using the general formulation of the previous
subsection, one finds that the first order perturbation, $Y^A_1(x)$, obeys
the equation
\be
{{{\p}^2{\cL}_0}\over{\p{\o}^A_{\a}\p{\o}^B_{\b}}}{\p}_{\a}{\p}_{\b}Y^B_1
+\left ({\o}^A_{\a}{{\p {\cL}_1}\over{\p{\o}^B_{\a}}}-{\d}_{AB}{\cL}_1\right )
{\p}_Bh(Y_0)=0
\label{49}\ee
The eq.(\ref{49}) represents a system of second order PDE with the
{\it constant}  coefficients
${{{\p}^2{\cL}_0}\over{\p{\o}^A_{\a}\p{\o}^B_{\b}}}$ and source terms
which are linear combinations of ${\p}_Bh(Y_0(x))$.
Now, using the expression (\ref{35}) or (\ref{39}) for ${\cL}_0(\Th )$,
one can show that
\be
{{{\p}^2{\cL}_0}\over{\p{\o}^A_{\a}{\o}^B_{\b}}}=
{\cL}_0\{ {\d}^{AB}{\O}^{\a\b}+{\o}^A_{\c}{\o}^B_{\d}
({\O}^{\a\c}{\O}^{\b\d}-{\O}^{\b\c}{\O}^{\a\d}-{\O}^{\a\b}{\O}^{\c\d}) \}
\label{50*}\ee
where ${\O}^{\a}_{\ \b}\= [(\bi +\bth )^{-1}]^{\a}_{\ \b}$.
Inserting this into the eq.(\ref{49}), we obtain
\be
I_{AB}{\O}^{\a\b}{\p}_{\a}{\p}_{\b}Y^B_1+{1\over{{\cL}_0}}
\left ({\o}^A_{\a}{{\p{\cL}_1}\over{\p{\o}^B_{\a}}}-{\d}^{AB}{\cL}_1 \right )
{\p}_Bh(Y_0)=0
\label{51*}\ee
where $I_{AB}\= {\d}_{AB}-{\O}^{\c\d}{\o}^A_{\c}{\o}^B_{\d}$.
It is easy to see, using eq.(\ref{2}), that the above equation
for $A=i,a,r$ (as defined in table (1)) is decomposed into the
three {\it uncoupled} equations
\bea
&&{\O}^{a\b}{\p}_{\a}{\p}_{\b} Y^i_1=0
\label{50}  \\
&&{\O}^{a\b}{\p}_{\a}{\p}_{\b} Z^a_1
-{{{\cL}_1}\over{{\cL}_0}}{\p}_ah(Y_0)=0
\label{51} \\
&&I_{rs}{\O}^{a\b}{\p}_{\a}{\p}_{\b}Y^s_1
+{1\over{{\cL}_0}}
\left ({\o}^r_{\a}{{\p {\cL}_1}\over{\p{\o}^s_{\a}}}-{\d}_{rs}{\cL}_1\right )
{\p}_sh(Y_0)=0
\label{52}\eea
where we have used the fact that ${\p}_ih=0$.
The operator ${\O}^{a\b}{\p}_{\a}{\p}_{\b}$ appearing above is indeed 
the D' Alembertian operator along the probe's worldvolume coordinates, as
can be seen by the eq.(\ref{12}), 
\be
{\O}^{a\b}{\p}_{\a}{\p}_{\b} =
-{1\over{1-v^2}}{\p}_0^{\ 2}+cos^2{\th}_1{\p}_1^{\ 2}+\ddd
+cos^2{\th}_p{\p}_p^{\ 2}
\label{52*}\ee
Thus the eqs.(\ref{50})-(\ref{52}) are written as the equations describing
the propagation of waves along the probe  with or without external sources.
Perturbations in the $y^i$ directions propagate as free waves,
as is expected by the homogeneity of the space along these coordinates,
while those in $z^a$ and $y^r$ directions
propagate as the forced oscillations.
The eq.(\ref{51}), 
describing the transverse oscillations, resembles a force equation 
 while eq.(\ref{52}), describing the longitudinal oscillations, is the
reminiscent of a torque equation.
In the equilibrium conditions, with $Y^A_1=0$, eqs.(\ref{51}),(\ref{52})
reduce to the eqs.(\ref{33}),(\ref{34}) for $n=1$ component.
These same conditions, for higher order $Y^A_n$'s, reproduce the
higher $n$ components of the eqs.(\ref{33}),(\ref{34}), and finally 
one recovers the eqs.(\ref{8}),(\ref{9}).  
As a result, our perturbative approach provides physical interpretations
for eqs.(\ref{8}),(\ref{9}) as the balancing conditions of the 
force and torque, respectively.

\section{Conclusion}
This paper categorizes several configurations of two arbitrary branes  
at angles which are 
derivable from the DBI+WZ action for p-branes. In using this dynamics for 
p-branes, we have implicitly assumed that all types of internal 
gauge fields of the branes as well as the background $B$ field are vanishing.
In this way all types of p-branes (e.g., NS-, D- and M-branes) are treated
in a similar manner by using the dynamics of DBI+WZ action.

Further, we have assumed that neither of the two branes is affected by the 
fields that originate from itself, at least when it is stretched as a flat
hypersurface (the BPS or no-force condition for single branes).
Thus, the WZ term contribute to the dynamics, when we deal with `similar'
branes carrying the same (p-form) charges, and it is vanishing when the   
the branes are `non-similar' carrying different charges. 
Under these assumptions, the analysis for both of these cases
reveals two types of configurations:
the marginal and the non-marginal ones. In the marginal case, we found 
that the only configuration of similar branes at angles is the one with
two angles in a subgroup of $SU(2)$, while the only one for non-similar  
branes is an orthogonal configuration obeying the ordinary intersection 
rule \cite{15}. In this case, no configuration with more than two angles 
can be found \cite{12}.  
In the non-marginal case, on the other hand, we saw that the only 
configuration with similar branes is the one with four angles in two 
independent subgroups of $SU(2)$, while the only one for non-similar
branes is an orthogonal configuration obeying an unusual rule of intersection,
previously identified as the localized branes intersection rule \cite{16}.
While, the whole analysis in this paper considers only two brane 
configurations, 
the $N$ brane configurations can also be undertaken by a similar
analysis, provided one knows the background fields for each  
$(N-1)$ of these branes.
Though, for marginal configurations, this does not seem to give additional  
information other than those for pair-wise intersections, it may be a useful
device for investigating about the non-marginal configurations made up
of several branes. 
It should be emphasized here that {\it marginal} multi-angle configurations, 
other than those stated in the above which 
all had been found in the context of the
classical supergravity solutions, no other marginal configuration can be
constructed from a set of {\it flat} p-branes. 
This is why the solutions of this 
kind had not been discovered in the supergravity solutions literature.
(This, however, does not prevent the possibility of having non-marginal 
solutions with several angles.) 
It is tempting to ask that whether one can form marginal 
multi-angle configurations by putting suitable curvatures on 
their worldvolumes. Of course, in such a case one has to define rigorously
the `marginality' property. In the case of asymptotic flat p-branes, we 
defined it as the stability  of the configuration  at {\it arbitrary }
separation of the two branes as measured in their {\it asymptotic} region.
 We have answered
the above  question, only partially , by perturbing 
the worldvolume of one of the branes when it is posed to the   
background of the other `unperturbed' brane, and find that the    
marginality requirement, in general, breaks the conditions for the flatness  
of the worldvolume. A general treatment requires propagating both of 
the branes and looking for the conditions that need to be asymptotically
flat.\\

{\Large{\bf Acknowledgements}}\\
I would like to thank H. Arfaei  for
fruitful discussions and  M.M. Sheikh-Jabbari for
careful reading of the paper and valuable suggestions.

\end{document}